# CAPTURE IN RESTRICTED FOUR BODY PROBLEM

Rosaev A.E.

OAO NPC NEDRA, Jaroslavl, Russia

A number of irregular moons of the Jovian planets have recently been discovered. Most adequate way of their origin is capture, but detailed mechanism is unknown. A few possibilities are discussed: collisions, gas drag, tidal destruction of a binary asteroid.

The capture process in restricted four body problem (RFBP) is researched. The interaction with regular satellite can be studied by this way as well as binary asteroid destruction in Hill sphere of planet. The energetic criteria of ballistic capture are studied and some numerical experiments are developed.

It is confirmed, that capture in four body problem is more probable on retrograde than on prograde orbit. In according with our results, encounter with regular satellites is more effective mechanism to create an irregular satellite population, than a binary asteroid flyby.

## 1. Introduction

All four large planets in the solar system possess irregular satellites, characterized by large, highly eccentric and/or highly inclined orbits. [1]. All these bodies were likely captured from heliocentric orbit, but detailed mechanism is unknown.

A distinctive feature of the irregular moons of the giant planets is their orbital grouping. Previously, the prograde and retrograde groups of irregular moons at Jupiter were believed to be groups of fragments produced by the disruption of two large moons. More recently, it has shown that the retrograde group has not one but probably four or more parent bodies. It is found that fragments were launched from two of the four identified parent moons, producing two clusters of irregular moons with members of each group having similar orbits [2]. Named the Ananke and Carme families, these two groups consist of seven and nine known member moons, respectively. The origin of this orbital clustering is unknown. Proper elements analyses, developed in [3], confirms presence of clusters. Current rates of collisions among satellites in the retrograde group are too low to explain them. Collisions with cometary impactors are even less likely. [2] Capturing irregular satellites via collisions between unbound objects can only account for about 0.1% of the observed population, hence can not be the source of irregular satellites. [4] Mutual collisions in action sphere maybe more effective mechanism [5-6].

The process of temporary capture is continued at present. Except well known example comet Shoemaker-Levi 9, f number of Jupiter family comets such as Oterma and Gehrels 3 make a rapid transition from heliocentric orbits outside the orbit of Jupiter to heliocentric orbits inside the orbit of Jupiter and vice versa. During this transition, the comet can be captured temporarily by Jupiter for one to several orbits around Jupiter. [7] Two asteroids 2001 QQ199 and 2004 AE9 and two comets P/LINEAR-Catalina and P/LINEAR are found to be quasi-satellites of Jupiter at present time [8]. We need to conclude, that capture is a very important process in Solar System dynamics. However, some details of capture are unknown for recent. In particular, the dependence of capture probability (and capture parameters) on target planet eccentricity is not clear.

There are a number of different capture modes are discussed. It has been thought [9-11] that the capture of irregular moons—with non-circular orbits—by giant planets occurs by a process in which they are first temporarily trapped by gravity inside the planet's Hill sphere. The capture of the moons is then made permanent by dissipative energy loss (for example, gas drag [9] or planetary growth [10]. But the observed distributions of orbital inclinations, which now include numerous newly discovered moons, cannot be explained using these models.

The known irregular satellites of the giant planets are dormant comet-like objects that reside on stable prograde and retrograde orbits in a realm where planetary perturbations are only slightly larger than solar ones. It is possible, that they may have been dynamically captured during a violent reshuffling event of the giant planets 3.9 billion years ago that led to the clearing of an enormous, 35M. disk of comet-like objects (i.e., the Nice model)..[11]

The role of chaos in capture dynamic is outlined in [12,13]. The key point is that incoming potential satellites get trapped in chaotic orbits close to "sticky" KAM tori in the neighbourhood of the planet, possibly for very long times, so that the chaotic layer largely dictates the final orbital properties of captured moons [12]

It is shown, that irregular satellites are captured in a thin spatial region where orbits are chaotic, and that the resulting orbit is either prograde or retrograde depending on the initial energy. Dissipation then switches these long-lived chaotic orbits into nearby regular (non-chaotic) zones from which escape is impossible. [13]

At present time, the model of capture object after destruction in action sphere, discussed by author in [14], becomes more popular. Triton inclined and circular orbit lies between a group of small inner prograde satellites and a number of exterior irregular satellites with both prograde and retrograde orbits. This unusual configuration has led to the belief that Triton originally orbited the Sun before being captured in orbit around Neptune. Agnor and Hamilton report, that a three-body gravitational encounter between a binary system (of $10^3$-kilometre-sized bodies) and Neptune is a far more likely explanation for Triton's capture. Our model predicts that Triton was on ce a member of a binary with a range of plausible characteristics, including ones similar to the Pluto–Charon pair. [15] Recent simulations of Triton's post-capture orbit could have followed either collisional or the recently-discussed three-body-interaction-based capture. [16]. The possible origin of the Martian moons Deimos and Phobos is capture at decomposition binary minor planet as it noted in [17]

It is natural to study capture of asteroid by planets in the restricted four body model. There are two abilities. We research case motion small body on boundary of planet's action sphere far from a planet's satellite. We apply direct energy calculation as it is defined in [18] (section 2). In section 3 some application and discussion of results are given.

## 2. Ballistic capture energy calculation

Let to consider restricted four body problem (RFBP). Main notations are: $m_1$ – main body (Sun), $m_2$ – secondary mass (Planet), $m_3$ – infinitesimal test particle, $m_4$ – mass, orbiting $m_2$ (Satellite), $f$ – is a gravity constant, $V_i$ – respective velocities, $\Delta_{ij}$ - distances between points. All bodies and velocities are in the same plane. Hamiltonian for this problem is:

$$H = 1/2\sum m_i V_i^2 - 1/2 f \sum_{j=0}^{n-1} \sum_{i=0}^{n-1} \frac{m_j m_i}{\Delta_{ij}} \qquad (1)$$

Let $m_1=1, V_1=0$.

$$H = 1/2 m_2 V_2^2 + 1/2 m_3 V_3^2 + 1/2 m_4 V_4^2 - f\frac{m_2}{\Delta_{12}} - f\frac{m_3}{\Delta_{13}} - f\frac{m_2 m_3}{\Delta_{23}} - f\frac{m_4}{\Delta_{14}} - f\frac{m_2 m_4}{\Delta_{24}} - f\frac{m_3 m_4}{\Delta_{34}} \qquad (2)$$

Denote $V_{23}$ - velocity of $m_3$ and $V_{24}$ - velocity of satellite $m_4$ relative the second primary:

$$V_3^2 = V_2^2 + V_{23}^2 - 2V_2 V_{23} \cos(V_2, V_{23})$$
$$V_4^2 = V_2^2 + V_{24}^2 - 2V_2 V_{24} \cos(V_2, V_{24})$$

$$H = 1/2(m_2 + m_3 + m_4)V_2^2 + 1/2 m_3 V_{23}^2 + 1/2 m_4 V_{24}^2 - m_3 V_2 V_{23} \cos(V_2, V_{23}) -$$
$$- m_4 V_2 V_{24} \cos(V_2, V_{24}) - f\frac{m_2}{\Delta_{12}} - f\frac{m_3}{\Delta_{13}} - f\frac{m_2 m_3}{\Delta_{23}} - f\frac{m_4}{\Delta_{14}} - f\frac{m_2 m_4}{\Delta_{24}} - f\frac{m_3 m_4}{\Delta_{34}}$$
(3)

or:

$$H = 1/2(-m_2 + m_3 + m_4)V_2^2 + 1/2 m_3 V_{23}^2 + 1/2 m_4 V_{24}^2 - m_3 V_2 V_{23} \cos(V_2, V_{23}) -$$
$$- m_4 V_2 V_{24} \cos(V_2, V_{24}) - f\frac{m_3}{\Delta_{13}} - f\frac{m_2 m_3}{\Delta_{23}} - f\frac{m_2 m_4}{r_{24}} - f\frac{m_3 m_4}{\Delta_{34}} - f\frac{m_4}{\Delta_{14}}$$
(4)

Denote $h_{23}$ - energy of $m_3$ relative the second primary.

$$h_{23} = 1/2 m_3 V_{23}^2 - f\frac{m_2 m_3}{\Delta_{23}}$$
(5)

*Definition. (Belbruno [18])* $P_3$ is ballistically captured at $P_2$ at time $t$ if the two body Kepler energy of $P_3$ with respect in $P_2$-centred inertial coordinates:

$$h_{23}(\mathbf{X}, \dot{\mathbf{X}}) = \frac{1}{2}\dot{\mathbf{X}}^2 - \frac{\mu}{|\mathbf{X}|} \leq 0, \quad 0 \leq \mu < 1/2$$

for a solution $\phi(t) = (\mathbf{X}, \dot{\mathbf{X}})$ of the elliptic restricted problem relative to $P_2$, $|\mathbf{X}|>0$

Let $\cos\beta \approx \pm\cos(V_2, V_{24})$, $\cos\alpha \approx \pm\cos(V_2, V_{23})$, $\cos\gamma \approx \cos(V_{23}, V_{24})$. Sign + is valid for retrograde, sign – for prograde orbits (Fig.1).

$$\Delta_{13} = \sqrt{r_{12}^2 + r_{23}^2 - 2r_{12}r_{23}\cos(\alpha)}, \Delta_{14} = \sqrt{r_{12}^2 + r_{24}^2 - 2r_{12}r_{24}\cos(\beta)}$$
$$\Delta_{34} = \sqrt{r_{24}^2 + r_{23}^2 - 2r_{23}r_{24}\cos\gamma}$$
(6)

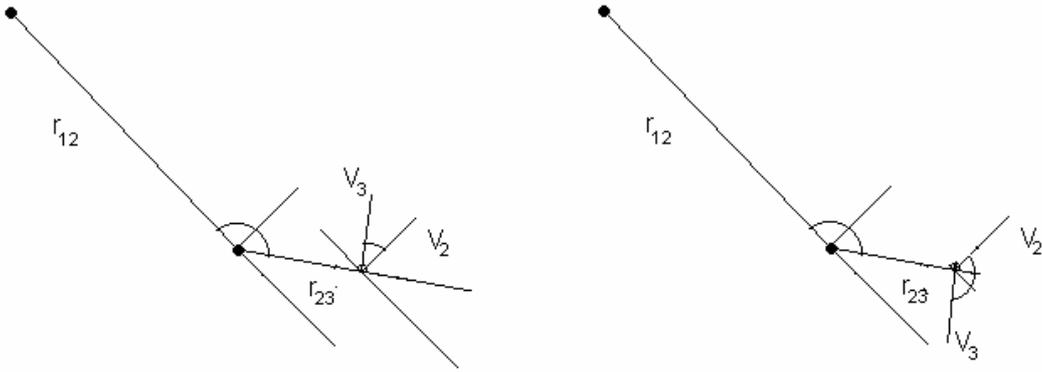

Fig 1. On definition of angles α and β for prograde (left) and retrograde (right) orbits.

At the action sphere enter:

$$h_{23} = H - (-m_2 + m_4 + m_3)V_2^2/2 \pm m_3 V_2 \sqrt{\frac{fm_2}{r_{23}}(1+e_{23})} \cos\alpha_b + \frac{fm_3}{\sqrt{r_{12}^2 + r_{23}^2 - 2r_{23}r_{12}\cos\alpha_b}} -$$
$$\pm m_4 V_2 \sqrt{\frac{fm_2}{r_{24}}(1+e_{24})} \cos\beta_b + \frac{fm_4}{\sqrt{r_{12}^2 + r_{24}^2 - 2r_{24}r_{12}\cos\beta_b}} - 1/2 m_4 V_{24}^2 + f\frac{m_2 m_4}{r_{24}} + f\frac{m_4 m_3}{\Delta_{34}}$$
(7)

At least, orbits of $m_1$, $m_2$ and $m_4$ are not change during encounter. Let $m_2$ move on fixed circular orbit around $m_1$ and $m_4$ move on fixed circular orbit around $m_2$ :

$$H' = H - (-m_2 + m_4 + m_3)V_2^2/2 - 1/2 m_4 V_{24}^2 + f\frac{m_2 m_4}{r_{24}} \approx const \tag{8}$$

approximately:

$$h_{23} = H' \pm m_3 V_2 \sqrt{\frac{m_2}{r_{23}}(1+e_{23})}\cos\alpha_b + \frac{m_3}{\sqrt{r_{12}^2 + r_{23}^2 - 2r_{23}r_{12}\cos\alpha_b}} \pm m_4 V_2 \sqrt{\frac{m_2}{r_{24}}}\cos\beta_b +$$
$$+ \frac{m_4}{\sqrt{r_{12}^2 + r_{24}^2 - 2r_{24}r_{12}\cos\beta_b}} + \frac{m_4 m_3}{\sqrt{r_{24}^2 + r_{23}^2 - 2r_{23}r_{24}\cos\gamma}} \tag{9}$$

After using Legendre expansion:

$$h_{23} = H' + m_3\left[\pm\sqrt{\frac{m_2}{r_{23}}} + \frac{r_{23}}{r_{12}}\right]\cos\alpha_f + m_3/2\left[\frac{r_{23}}{r_{12}}\right]^2(3\cos^2\alpha_f - 1) +$$
$$+ m_4\left[\pm\sqrt{\frac{m_2}{r_{24}}} + \left[\frac{r_{24}}{r_{12}}\right]\right]\cos\beta_f + m_4/2\left[\frac{r_{24}}{r_{12}}\right]^2(3\cos^2\beta_f - 1) + \frac{m_4 m_3}{\Delta_{34}} \tag{10}$$

Or, more precisely:

$$h_{23} = H' + \left[\pm m_3\sqrt{\frac{m_2}{r_{23}}(1+e_{23})} + \frac{m_3 r_{23}}{r_{12}}\right]\cos\alpha_f + m_3/2\left[\frac{r_{23}}{r_{12}}\right]^2(3\cos^2\alpha_f - 1) +$$
$$+ m_3/2\left[\frac{r_{24}}{r_{12}}\right]^2(3\cos^2\beta_f - 1) + m_3/2\left[\frac{r_{23}}{r_{12}}\right]^3(5\cos^3\alpha_f - 3\cos\alpha_f) + m_4\left[\pm\sqrt{\frac{m_2}{2r_{24}}} + \left[\frac{r_{24}}{r_{12}}\right]\right]\cos\beta_f +$$
$$+ m_4/2\left[\frac{r_{24}}{r_{12}}\right]^3(5\cos^3\beta_f - 3\cos\beta_f) + \frac{m_4 m_3}{\Delta_{34}}$$

## 3. Application to Jupiter system

There are two different manifolds of orbits of encounter, which can provide transfer to satellite orbit. First kind of them – close-to-parabolic trajectories, passed deeply inside in action sphere of planets. In spite of formally small amount of energy change needs to capture, actually is necessary to take into account, that apocenter of final orbit of capture need to be inside of action sphere. This fact significantly increase required energy change.

Second case (fig.) of possible orbit of capture is elliptic (e<1) orbits on boundary of action sphere (Hill radius). At these conditions, small eccentricity is not guarantee fact of capture. Motion seems to be very chaotic, however, small change of energy (small perturbations) can completely change character of orbits. A few examples of such class orbit of capture are given in a previous author's numeric investigations [19]. This class seems to be more perspective for real capture.

Here we consider second case. The perturbing satellite is far from test particle $m_3$. We can neglect last term. The necessary conditions for capture are:

$$h_{23b} = m_3 \left[ \sqrt{\frac{m_2}{r_{23}}(1+e_{23b})} \cos\alpha_b \right] + m_3 \frac{r_{23}}{r_{12}}(\cos\alpha_b) + m_4 \left[ \sqrt{\frac{m_2}{r_{24}}} + \frac{r_{24}}{r_{12}} \right](\cos\beta_b) > 0$$
$$h_{23f} = m_3 \left[ \sqrt{\frac{m_2}{r_{23}}(1+e_{23f})} \cos\alpha_f \right] + m_3 \frac{r_{23}}{r_{12}}(\cos\alpha_f) + m_4 \left[ \sqrt{\frac{m_2}{r_{24}}} + \frac{r_{24}}{r_{12}} \right](\cos\beta_f) < 0$$

(11)

For Jupiter:

$$h_{23b} = \left[ \sqrt{\frac{0.001}{0.113}(1+e_{23b})} \cos\alpha_b \right] + 0.113(\cos\alpha_b) + \frac{m_4}{m_3}\left[ \sqrt{\frac{0.001}{0.012}} + 0.012 \right](\cos\beta_b) > 0$$
$$h_{23f} = \left[ \sqrt{\frac{0.001}{0.113}(1+e_{23f})} \cos\alpha_f \right] + 0.113(\cos\alpha_f) + \frac{m_4}{m_3}\left[ \sqrt{\frac{0.001}{0.012}} + 0.012 \right](\cos\beta_f) < 0$$

(12)

Rough estimation, based on Kepler's law give: $\beta \approx 27\alpha$. In energetic calculation, eccentricity change cannot be properly estimated. On the other hand, at action sphere boundary, e<1 not guarantee capture. So we put e=0.5 fixed. Similar character of h(f) dependence true for wide range of eccentricity variations 0.1<e<1.95.

In case three body problem, we have only one area of zero energy when capture is possible, mainly at large value of true anomaly (Fig.2), i.e. large arc of orbit in planets action sphere. It makes probability of capture small. In four body case, phase space becomes much more complex: this fact significantly increase capture possibility, in particular, at small part of orbit passed in action sphere (Fig.3). Effect of close-to planet satellite is similar tidal badge in old model tidal capture, but seems more effective.

There is an important difference between case prograde and retrograde orbits. For case prograde orbital motion we have a symmetric dependence on true anomaly and range of true anomaly, where energy is negative is independent on satellite mass (fig.4). It means that distant satellite occur small effect to capture. In contrary, in case capture on retrograde orbit, dependence on true anomaly is asymmetric and interval of negative energy significantly increases with mass satellite increasing (Fig.5). It means that of retrograde orbits, capture more probable than for prograde ones and relatively large objects can be captured on retrograde objects. At capture on retrograde orbit, presence of regular satellite change time reversibility of trajectory more effectively, than for prograde ones.

Finally, we can conclude, that indirect perturbation of distant satellite can strongly increase probability to capture small body onto retrograde orbit. Of course, stability of retrograde orbits is higher than prograde. But here we have show, that fact of capture is much more possible for retrograde satellites. It explains the well known observation evidence, that most of distant satellites of giant planets are retrograde. Note, that it is not take place for Neptune system, where largest satellite Triton moved in retrograde orbit.

### Conclusions

The capture process in restricted four body problem (RFBP) is researched. There are two kinds of capture possibility. We consider motion infinitesimal particle at a boundary of action sphere of planet, far from a perturbing satellite.

The expression for test particle energy is derived. The map in a phase space in RFBP model is much more complex, than in a three body problem. Possible it gives new abilities for energy dissipation.

A significant difference between case prograde and retrograde orbits is found, maybe capture in four body problem is more probable on retrograde than on prograde orbit.

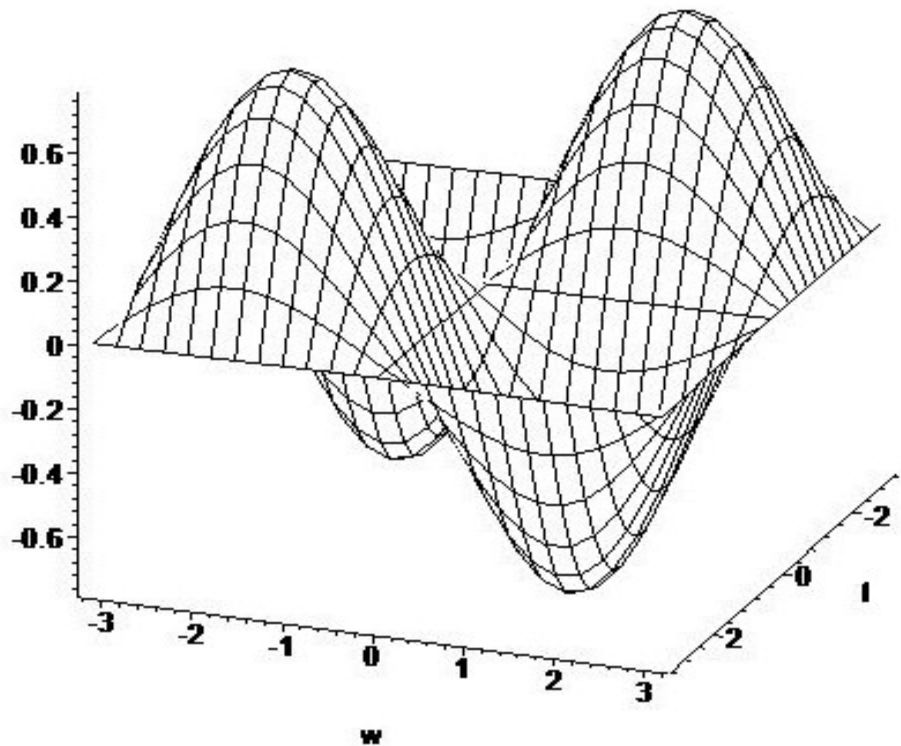

Fig. 2. Energy of capture $h_{23}$ on pericentre longitude w and true anomaly alpha dependence (three body problem).

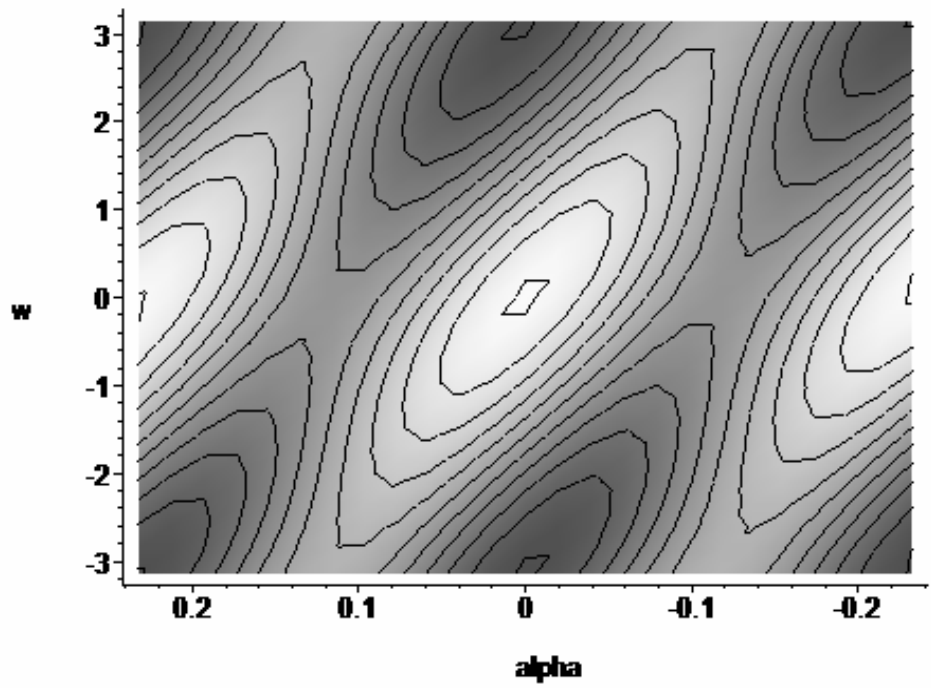

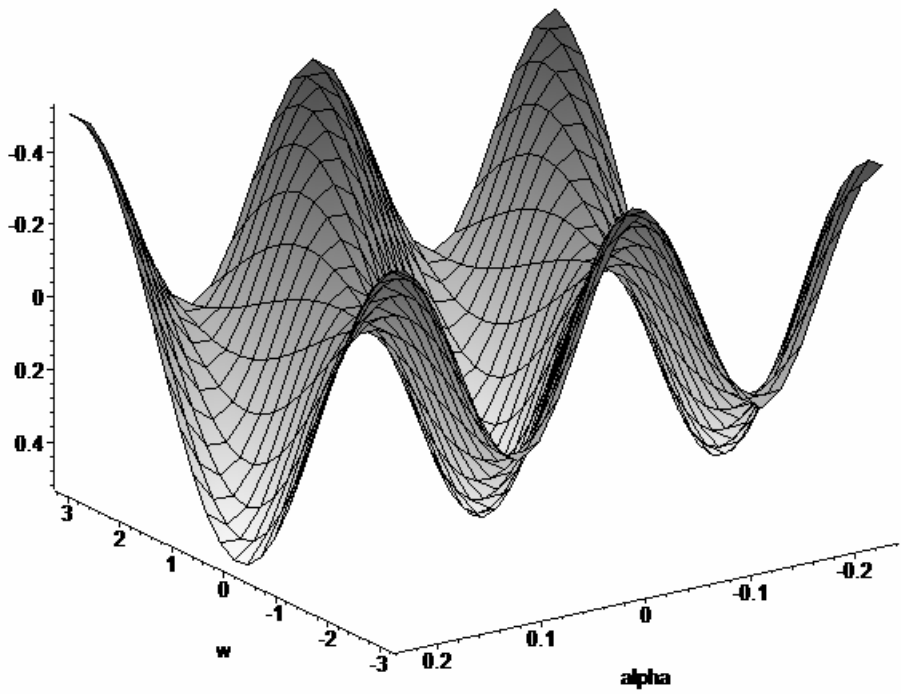

Fig. 3 Energy of capture $h_{23}$ on pericentre longitude w and true anomaly alpha dependence. Dark area respect negative energy. (four body problem)

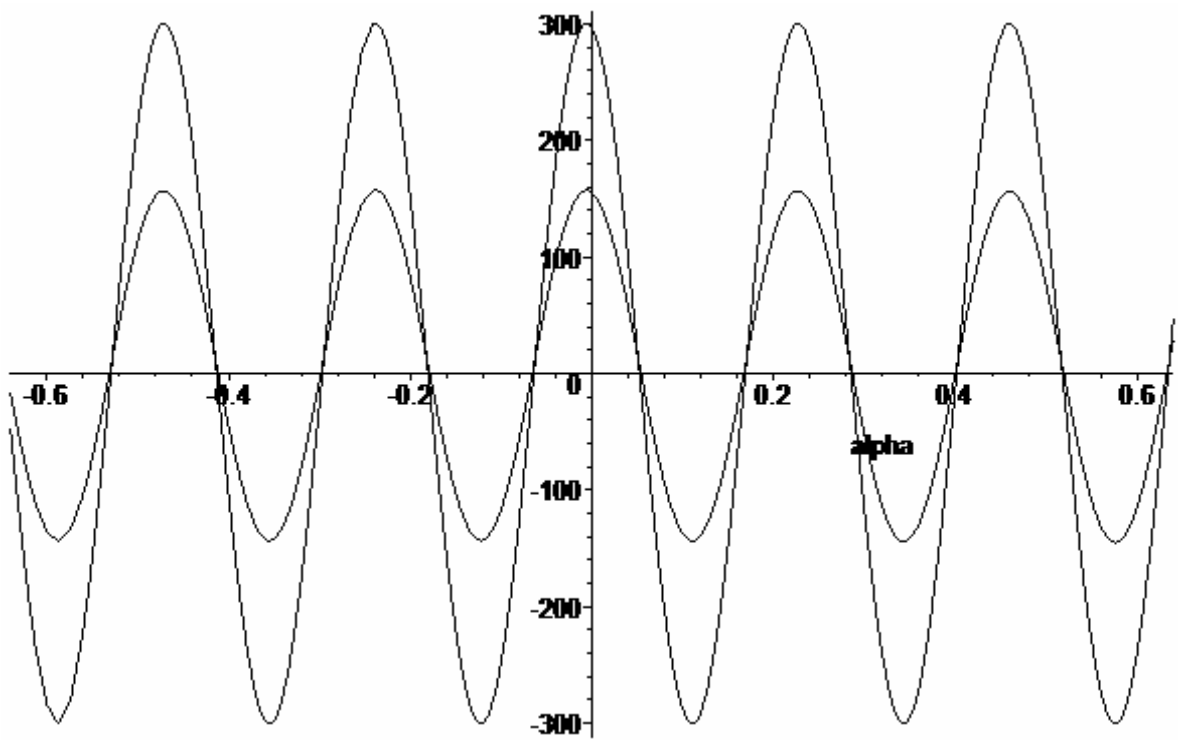

Fig. 4. Test particle relative energy ($h_{23}$) of on phase $f$ dependence.
Case of prograde motion.

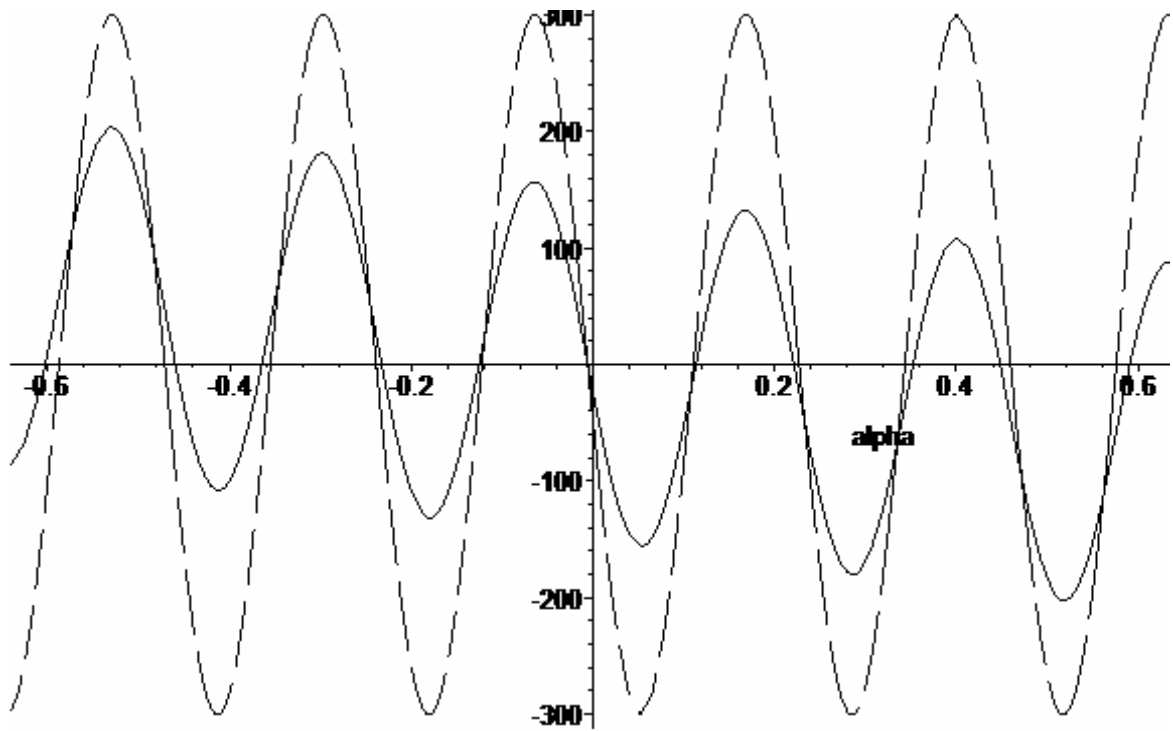

Fig.5 Test particle relative energy ($h_{23}$) of on phase $f$ dependence.
Case of retrograde motion. Solid line: $m_4/m_3$=1, dashed line: $m_4/m_3$=1000.